# THE MOIST GREENHOUSE IS SENSITIVE TO STRATOSPHERIC TEMPERATURE

Ramses M. Ramirez[i]

[i]Earth-Life Science Institute, Tokyo, Japan

email: rramirez@elsi.jp

## INTRODUCTION

Recent 3-D model simulations (Fujii et al. 2017; Kopparapu et al. 2017) (hereafter, F2017 and K2017) suggest that tidally-locked M-star planets at the inner edge of the habitable zone (HZ) undergo a moist greenhouse at significantly lower mean surface temperatures (~280 – 300 K) than the ~340 K predicted in 1-D models (e.g. Kasting et al., 1993). This moist greenhouse occurs once stratospheric $H_2O$ concentrations exceed ~$3\times10^{-3}$, when the cold trap is wet enough for atmospheric $H_2O$ to efficiently dissociate so that an amount of water equal to that in Earth's oceans can escape to space on ~4.5 Gyr timescales.

F2017 attribute these differences to strong near-infrared absorption (NIR) that triggers upward motions that increase upper atmospheric $H_2O$ concentrations more rapidly than predicted by 1-D models. F2017 also (correctly) argue that dynamical effects are very important to accurately characterize climates, which is something that 1-D models do not include explicitly (although this is often parameterized implicitly). Although true, this is not the main reason for the apparent discrepancy between 1-D and 3-D model results for calculated moist greenhouse thresholds on M-star planets.

It is known that the moist greenhouse mean surface temperature threshold is very sensitive to the stratospheric temperature (e.g. Kasting et al., 1993). In 1-D calculations (in lieu of knowing anything about the exoplanetary atmosphere), a 1-bar fully-saturated $N_2$ atmosphere on a planet with an Earth-like surface water inventory and (most importantly) a constant stratospheric temperature of 200 K (above the cold trap) is traditionally assumed. However, the 3-D simulations in both F2017 and K2017 exhibit upper atmospheric and stratospheric temperatures well above 200 K as the moist greenhouse threshold is approached on their M-star planets. This note is a reminder of the importance of stratospheric temperatures for this particular calculation.

## METHODS

Here, I use my single-column climate model (e.g. Ramirez and Kaltenegger, 2017) to construct idealized cases that illustrate how the moist greenhouse surface temperature threshold changes as stratospheric temperature is increased. Atmospheres are fully-saturated and stratospheric temperature is assumed constant.

## RESULTS

At a stratospheric temperature ($T_{strat}$) of 200 K, the moist greenhouse is triggered at a mean surface temperature of ~340 K (as expected) (Figure 1). However, it commences at significantly lower mean surface temperatures as $T_{strat}$ is increased. At a $T_{strat}$ of (220, 243, 260) K, the moist greenhouse is triggered at surface temperatures of ~ (330, 309, 280) K (Figure). The latter two cases are consistent with the bulk trends of K2017 and F2017.

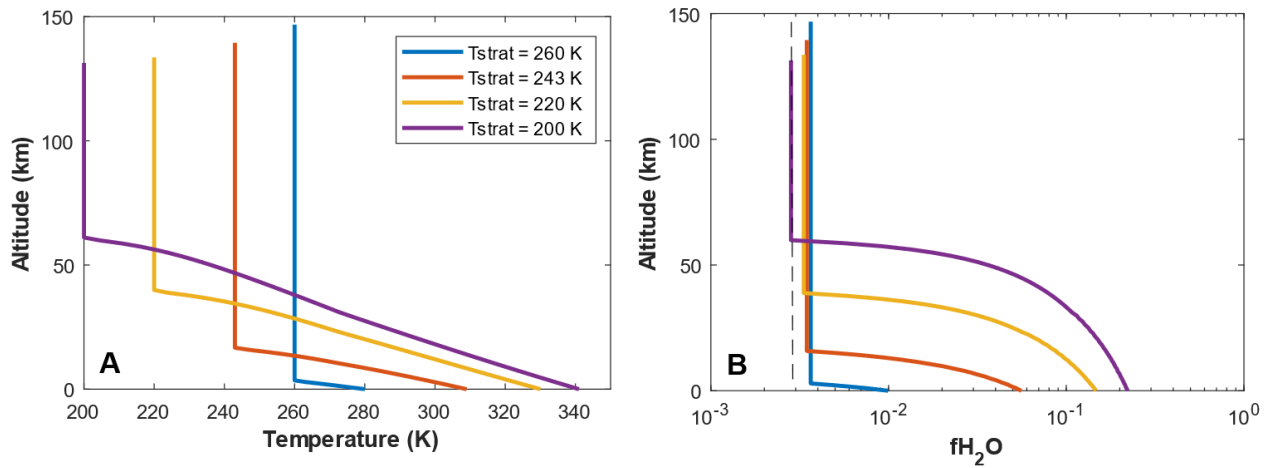

**Figure 1:** Vertical profiles of a) temperature and b) water vapor mixing ratio ($fH_2O$) for different stratospheric temperature assumptions (200, 220, 243 and 260 K). A moist greenhouse is triggered at stratospheric $fH_2O > \sim 3 \times 10^{-3}$ (dashed).

# DISCUSSION

Thus, at higher $T_{strat}$, the atmosphere holds more water vapor at higher altitudes above the cold trap (consistent with the Clausius-Clapeyron equation), which accelerates the transition to a moist greenhouse. In 3-D models, this is represented by enhanced upward transport of water vapor.

Thus, most discrepancies with the moist greenhouse computation are likely not due to perceived deficiencies in 1-D (or 3-D) models, but it helps to understand why certain assumptions in 1-D models were made. The $T_{strat} = 200$ K assumption is convenient for HZ calculations because we do not know whether habitable M-star planets with relatively low $T_{strat}$ do not exist, perhaps under different atmospheric and planetary assumptions.

Nevertheless, as $T_{strat}$ is directly correlated to NIR absorption, which is generally more intense on M-star planets, it would be possible to compute $T_{strat}$ as a function of star type using basic physical principles, and in turn, determine new moist greenhouse thresholds for such planets. I should note that 1-D and 3-D models are often consistent with one another when an appropriate methodology is employed.

Unfortunately, on those M-star planets in which $T_{strat}$ is high, their habitability is in doubt, however. If such planets are currently undergoing moist greenhouses they would likely be desiccated. Unless originally located far away, these worlds would have lost *tens - hundreds* of Earth oceans during the superluminous pre-main-sequence stellar phase, which can last over 2 Gyr (e.g. Ramirez and Kaltenegger, 2014; Luger and Barnes, 2015). Also, because M-star luminosity varies little during the main-sequence, these planets may have been in this moist greenhouse state for many Gyr. Thus, unless their water inventories (by mass) exceed a few tens % of their planetary mass (e.g. Levi et al. 2017), they may be dry. Assuming such desert worlds are habitable, spectral features may be too weak to be detected.

# CONCLUSION

Both 1-D and 3-D climate modelers should collaborate to understand both the common points and differences in our approaches. We can learn a lot from one another on the moist greenhouse problem among others.

**ACKNOWLEDGEMENTS:** I acknowledge enlightening discussions regarding the implications of GCM results with Eric Wolf, Yuka Fujii, and Tony Del Genio. I also acknowledge support from the Earth-Life Science Institute.